\documentclass[letterpaper]{article}
\usepackage{aaai25}
\usepackage{times}
\usepackage{helvet}
\usepackage{courier}
\usepackage[hyphens]{url}
\usepackage{graphicx}
\usepackage{natbib}
\usepackage{caption}
\nocopyright
\usepackage{booktabs}

\usepackage{algorithm}
\usepackage{algorithmic}
\usepackage{newfloat}
\usepackage{listings}
\DeclareCaptionStyle{ruled}{labelfont=normalfont,labelsep=colon,strut=off} 
\floatstyle{ruled}
\newfloat{listing}{tb}{lst}{}
\floatname{listing}{Listing}
\title{Race Discrimination in Internet Advertising:  Evidence From a Field Experiment}

\author {
    Neil K. R. Sehgal\textsuperscript{\rm 1}, 
    Dan Svirsky\textsuperscript{\rm 2}
}
\affiliations {
    \textsuperscript{\rm 1}Department of Computer \& Information Science, University of Pennsylvania; Philadelphia, PA, USA\\
    \textsuperscript{\rm 2}Harvard Business School; Boston, MA, USA. Current affiliation: Uber Technologies, Inc.; Boston, MA, USA\\
    neilsehgal99@gmail.com, dan.svirsky@gmail.com
}

\usepackage{bibentry}

\begin{document}

\maketitle

\begin{abstract}
We present the results of an experiment documenting racial bias on Meta’s Advertising Platform in Brazil and the United States. We find that darker skin complexions are penalized, leading to real economic consequences. For every \$1,000 an advertiser spends on ads with models with light-skin complexions, that advertiser would have to spend \$1,159 to achieve the same level of engagement using photos of  darker skin complexion models. Meta’s budget optimization tool reinforces these viewer biases. When pictures of models with light and dark complexions are allocated a shared budget, Meta funnels roughly 64\% of the budget towards photos featuring lighter skin complexions.\end{abstract}

%

\section{Introduction}
This paper seeks to measure racial discrimination in an online advertising platform, as well as understand the mechanisms underpinning said discrimination and the economic costs it imposes on historically marginalized groups.

Many important marketplaces that used to operate in physical spaces have moved online. Ads shift from billboards to timelines. Wedding photographers advertise on Instagram instead of the Yellow Pages. These changes are economically important: more than half of the \$300 billion spent on U.S. advertising in 2022 was spent on just three online platforms: Amazon, Google, and Meta \cite{mcgee2022meta}. As these markets have moved online, research documenting race discrimination has continued to find that racial biases have an impact on important outcomes, online and off. Discrimination by ethnicity has been well-documented in marketplaces from the market for credit, goods, labor, short-term rentals, crime enforcement, and housing \cite{pope2010picture, doleac2013visible, pager2003mark, agan2017ban, edelman2017racial, horrace2016dark, hanson2011landlords}.

This paper builds on that literature with an experiment documenting the role of race discrimination on Meta’s advertising platform in Brazil and the United States. We measure whether photographs of people with darker complexions garner less engagement and how any disparities translate into monetary penalties. We further measure whether Meta’s optimization algorithms contribute to discrimination or mitigate it. In the experiment, we run advertisements for wedding photographers using photographs of models that vary in their skin complexion. We use a 2x2 design: we compare pairs of photographs that are similar in every way except skin complexion to get a baseline measure of differences in engagement. Then we run advertisements using photographs from the same pairing, but zoomed in and cropped so as to remove non-race related features (e.g., details of a dress or a flower arrangement) and make the skin complexion of the models a more salient part of the ad. We find that when advertisements highlight subjects with darker skin, they receive 10.39\% fewer likes. This difference has economic significance: advertisers must spend 11.59\% more per photo to garner the same engagement for a picture highlighting a person of color. Observational data on the demographics of the audience suggest that Meta is actively showing the different ads to different types of users, but we do not find evidence that these under-the-hood decisions make the treatment effect stronger.

We also find that Meta’s budget optimization tool, while on its face neutral, exacerbates discrimination by reflecting user bias. Meta offers advertisers a tool to optimize budget decisions by spending more money on ads that get more engagement. In our experiment, we find that the platform’s budget optimization tool funnels roughly 64\% of advertising dollars towards pictures of models with light skin complexions.

Our findings have implications for the legal regulation of online markets, because they show how facially neutral algorithms can reinforce user biases to make racial disparities worse. Our findings also contribute to the social science literature on discrimination by showing that racial disparities exist even in settings where statistical discrimination is a less natural explanation, compared to straightforward taste-based animus.

\section{Methods}

We conduct an experiment to measure whether ads for wedding photographers featuring people with darker complexions get fewer likes than ads for wedding photographers featuring people with lighter complexions. We further test whether Meta’s optimization tools affect any underlying disparities we find. We also measure whether, as a result of any disparity, photos featuring people with darker complexions require higher advertising costs to garner similar levels of audience engagement. We pre-registered the experiment on osf.io, the Center for Open Science’s repository. All code and data is available in this repository as well.

\subsection{Image Selection and 2x2 Design}
Ideally, we would take two photos that are identical in everything except the skin color of the subject and measure whether changing the skin color leads to fewer likes. One could compare ads that have models who look similar, in a similar pose and context, but who vary by complexion. Or, one could take an ad with someone with light skin and make the skin complexion look darker, or someone with dark skin and make the skin complexion look lighter, using photo editing software. Then the research question is straightforward -- do two otherwise similar pictures have different engagement levels when the skin color changes? 

But such an approach is imperfect because it is impossible to only change skin color without affecting other attributes of the photograph. Shadows, lighting, contrast, hues, saturation -- all of these traits are important to the aesthetic value of a photograph and are hard to control or measure by the experimenter. Pictures, after all, are worth an aphoristic thousand words. And using any one pair of pictures raises external validity concerns -- would such findings apply to other pictures? 

We address this challenge using the 2x2 experimental design illustrated in Figure 1. Consider the two pictures in the two right most columns -- a zoomed out picture of a bride and groom embracing. Simply measuring the difference in Likes for the darker skin versus the lighter skin suffers from the problem described above: unobserved or unmeasured differences between the pictures that correlate with skin tone. 

Our measure of racial discrimination is different. Given a baseline difference in Likes for two similar pictures with different skin complexions, we ask whether there is a racial penalty for zoomed-in versions of these same photographs, where the skin takes up a bigger portion of the picture and is therefore more salient, relative to the baseline racial difference of the zoomed-out versions of these same photographs.

\begin{figure*}[t]
\centering
\includegraphics[width=0.8\textwidth]{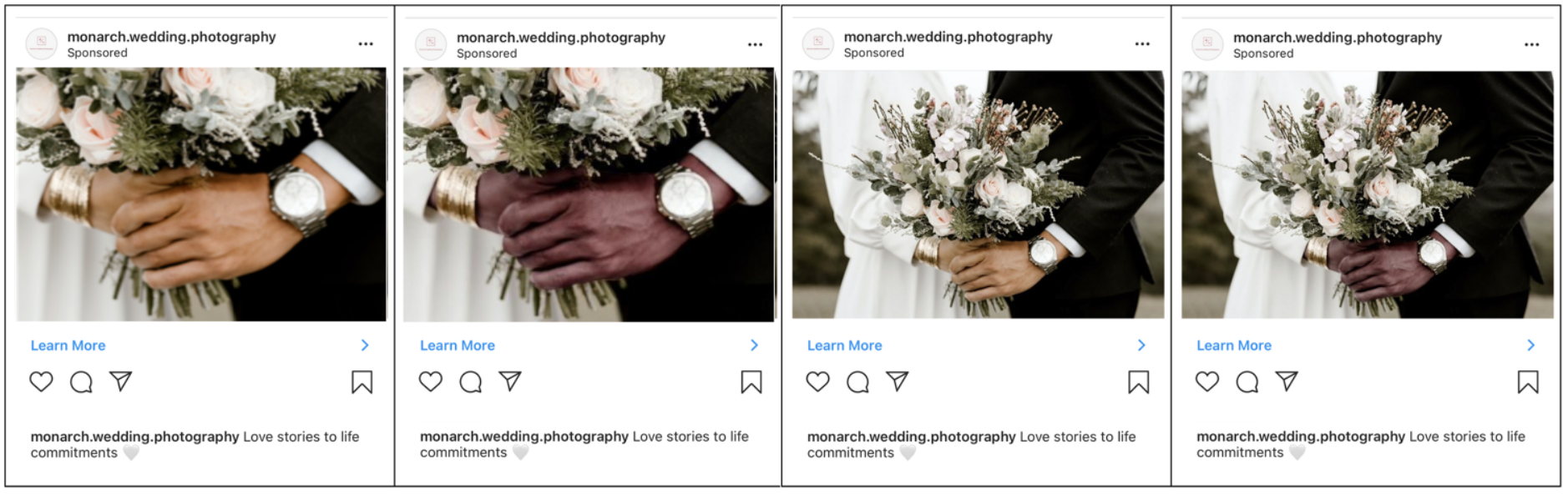} 
\caption{Illustration of the 2x2 experimental design. In the experiment, we compare two similar pictures where skin complexion differs to measure differences in how many users “Like” each ad on Instagram. We conduct a 2x2 design by then zooming in on each picture in a way that makes the skin more salient and then measuring whether this creates any disparities in “Like” rates between the two pictures, after controlling for any baseline differences in “Like” rates.}
\label{fig1}
\end{figure*}

The dependent variable of interest is the difference in the proportion of likes for the two ads, when we present a zoomed-in photograph where skin complexion is more salient, relative to the difference between the baseline (zoomed-out) pictures. 

Specifically, let $P_{L}$ be equal to the number of likes for a photograph of a model with a lighter complexion divided by the total number of people who viewed that photograph, $P_D$ be the same for a photograph of a model with a dark complexion, $P_{LZ}$ be the same for the zoomed-in photograph of the model with a light complexion, and $P_{DZ}$ be the same for a zoomed-in photograph of the model with a dark complexion. The change in ${Likes / Views}$ when an ad zooms in and makes the skin in a picture more salient can be expressed, for people with light skin, as follows:

\begin{equation}
\left(P_{L Z}-P_L\right)
\label{eq:one}
\end{equation}

And similarly for people with darker complexions. 

We are most interested in whether the following equation holds:

\begin{equation}
\left(P_{L Z}-P_L\right)-\left(P_{D Z}-P_D\right)=0
\label{eq:two}
\end{equation}

If equation \ref{eq:one} does not hold, and if the expression on the left is positive, then this is evidence that making the skin more salient penalizes ads with people with darker complexions. If the expression on the left is negative, then this is evidence that making the skin more salient penalizes ads with people with lighter complexions. If the expression holds, then this is evidence of a null effect.

Another reasonable measure of discrimination is in equation \ref{eq:three}:

\begin{equation}
\left(P_{L Z}+P_L\right)-\left(P_{D Z}+P_D\right)=0
\label{eq:three}
\end{equation}

This equation tests whether the total engagement rate for photos with models with light complexions is equal to the total engagement rate for photos with models with dark complexions. If the expression is greater than zero, that suggests a bonus when the photo features lighter complexions.

We consider Equation \ref{eq:two} a cleaner test, as it measures whether darker complexions see a penalty when skin is more salient relative to a baseline that tests underlying differences in the pictures themselves. For that reason, we pre-registered Equation \ref{eq:two} as the primary outcome of interest. Nonetheless, Equation \ref{eq:three} may be of independent interest, and we report both.

As noted above, the best approach for this experiment is not obvious ex ante -- similar photographs of models with different complexions? An identical photograph with the complexion changed in Adobe Photoshop? We address this challenge with an all-of-the-above approach. We run the experiment six times, with six pairs of photographs. In Pairs 1 and 2, the photographs are of different subjects but holding the same pose. In Pairs 3 and 4, the photos are identical, but the skin complexion in an original photo was made darker with Adobe Photoshop. In Pairs 5 and 6, the photos are identical, but the skin complexion in an original photo was made lighter with Adobe Photoshop.

Using six sets of photos also helps to address the second challenge described above -- external validity. If we find a penalty for skin complexion for one pair of photographs, but not for another, then this is not clear evidence of discrimination. 

Figure 2 summarizes the six tests we run.

\begin{figure*}[t]
\centering
\includegraphics[width=0.55\textwidth]{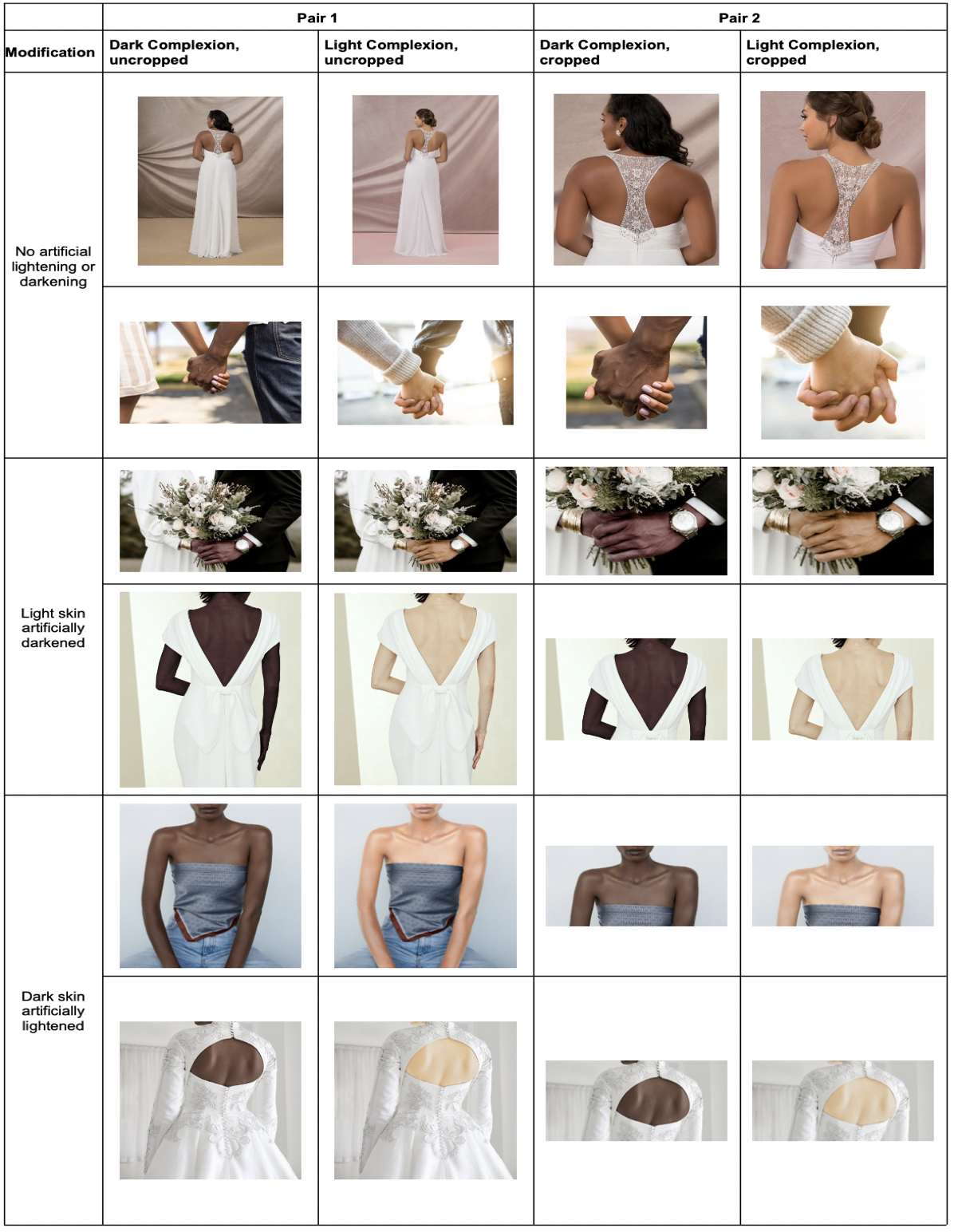} 
\caption{Images tested. We test six sets of images in the experiment, running four ads for each of the six sets. Each set has four pictures: two similar pictures where the skin complexion differs, then two identical pictures, but zoomed in. In two of the sets, we find pictures that look similar but with different models with different skin complexions. In two of the sets, we take one picture and use Adobe Photoshop to artificially make the skin complexion look lighter. In the remaining two sets, we take one picture and use Adobe Photoshop to artificially make the skin complexion look darker. }
\label{fig2}
\end{figure*}

\subsection{Meta Advertising Platform}

Our experiment ran on Meta’s advertising platform, which we describe here.

Advertisers on Meta are presented with an advertising dashboard that allows for customization of an ad campaign. Meta lets advertisers select their own criteria for targeting ads, albeit with some limitations following a civil rights lawsuit. An advertiser can aim their ad at age groups, by geography, by interests (like sports or wedding photography), or by whether a person is similar to an existing group of Meta users. Meta determines user interests through multiple factors including past page and advertisement engagement, demographics, and network connection speed. When creating an ad, Meta offers 11 unique Campaign Objectives for users to choose. Based on the selected objective, Meta serves ads to different audiences based on who it believes is most likely to take a desired action. Examples of objectives include audience reach, which tries to show the ad to as many people as possible, and audience engagement, which tries to maximize the number of Likes, comments, and shares for an ad.

Meta exercises significant control over who sees each advertisement in ways that are opaque. In theory, an advertiser could simply give Meta a list of phone numbers and ask the platform to randomly select a subset of this audience to view the ad. But recent research – and our results below -- suggests Meta does not always do this. Past research shows that Meta uses its user data -- a person’s likes, friends, interests, and so on -- to target ads based on the content of those ads, effectively choosing the audience reach based on who it thinks will respond to the ad \cite{ali2019discrimination}. 

\subsection{Advertisement Creation}
We run an ad on Instagram for each of the twenty-four images using the Meta Ad Manager interface. We focus on the Engagement objective, which targets users most likely to engage with an ad through follows, comments, shares, or likes. We set the audience to Instagram users in the United States, age 18 years or older, and with an interest in wedding photography. This specific audience yielded an eligible audience of 9.6 million Instagram users. For comparison, an audience of Instagram users in the United States, age 18 years or older, but with no specified interests yielded an eligible audience of 130 million Instagram users.

We place ads exclusively on the Instagram Feed. Advertisements in the Instagram Feed are identical to normal posts, except for a small “Sponsored” disclosure and a clickable “Learn More” link, directing viewers towards our ad account’s profile page. Within each group, the four ads possess an identical account name and caption, only differing by image. Given the large eligible audience of millions of users, it is unlikely any individual Instagram user would have seen more than a single ad from one grouping. Figure 1 shows an example of an advertisement set. Based on data from pilot testing for our selected audience, an ad is viewed by 1000 Instagram users for every \$15.86 spent. We budgeted each ad to be shown to 1160 users (\$18.39) over 24 hours based on a power calculation with Beta set to 0.2 and Alpha set to 0.05 for an minimum detectable effect of a five percentage point difference in likes between two pictures (also based on pilot testing). This study is not considered human subjects research because no identifiable or private information was collected from people who viewed the ads.

\section{Results}

\subsection{Main Experiment Results: Who Saw the Ads}

In total, 34,419 Instagram users viewed one of the 24 advertisements. Users responded with a Like in 7,530 cases for an average like/ad view of 0.22. Summary statistics for the demographics of the advertisement viewers are listed in Table 1. A majority of the viewer population is female and age 18-24.

\begin{table}[ht]
\centering
\resizebox{\columnwidth}{!}{%
\begin{tabular}{@{}llccccc@{}}
\toprule
\multicolumn{2}{l}{\textbf{Category}} & \textbf{All Ads} & \textbf{Dark Cropped} & \textbf{Dark Uncropped} & \textbf{Light Cropped} & \textbf{Light Uncropped} \\
\midrule
\textbf{Gender} & Female & 23,462 (68.2\%) & 3,776 (52.5\%) & 8,383 (80.3\%) & 2,482 (39.2\%) & 8,821 (84.4\%) \\
              & Male   & 10,697 (31.1\%) & 3,356 (46.7\%) & 1,987 (19.0\%) & 3,796 (59.9\%) & 1,558 (14.9\%) \\
              & Unknown & 260 (0.8\%) & 57 (0.8\%) & 76 (0.7\%) & 57 (0.9\%) & 70 (0.7\%) \\
\midrule
\textbf{Age}   & 13-17 & 1 (0\%) & 0 (0\%) & 0 (0\%) & 0 (0\%) & 1 (0\%) \\
              & 18-24 & 17,222 (50.04\%) & 2,955 (41.1\%) & 5,701 (54.6\%) & 2,099 (33.1\%) & 6,467 (61.9\%) \\
              & 25-34 & 8,236 (23.93\%) & 1,791 (24.9\%) & 2,657 (25.4\%) & 1,479 (23.3\%) & 2,309 (22.1\%) \\
              & 35-44 & 3,925 (11.4\%) & 1,094 (15.2\%) & 965 (9.2\%) & 1,125 (17.8\%) & 741 (7.1\%) \\
              & 45-54 & 2,603 (7.56\%) & 706 (9.8\%) & 562 (5.4\%) & 881 (13.9\%) & 454 (4.3\%) \\
              & 55-64 & 1,606 (4.67\%) & 439 (6.1\%) & 360 (3.4\%) & 519 (8.2\%) & 288 (2.8\%) \\
              & 65+ & 826 (2.4\%) & 204 (2.8\%) & 201 (1.9\%) & 232 (3.7\%) & 189 (1.8\%) \\
\bottomrule
\end{tabular}%
}
\caption{Breakdown of Characteristics by Treatment Status. This table shows the demographics of ad viewers across all advertising campaigns and within the four types of advertisements (cropped/uncropped with dark/light complexion).}
\end{table}

The most striking result is that Meta’s advertising platform is making different under-the-hood decisions about which groups to serve each ad to. This is not a novel finding. Past research has documented that even when Meta’s Advertising Platform is given a list of randomly chosen American phone numbers and all ad targeting is turned off, Meta still directs, for example, makeup ads to women \cite{ali2019discrimination}. Table 1 replicates this finding. It is not publicly known how Meta’s algorithms make these choices. For purposes of this paper, it means that any treatment effect we find -- any racial bias -- could be driven by user choices, by Meta’s decision about who to serve ads to, or by a combination of the two.

\subsection{Main Experiment Results: Treatment Effect}

We find evidence of a penalty for pictures of models with darker complexions. Table 2 presents the results of a regression measuring the treatment effect. Column 1 is a linear regression regressing whether the ad viewer Liked the picture on three variables: whether the picture has a model with darker skin complexion, whether the picture is cropped, and the interaction of these two. The variable of interest is the interaction between the complexion featured in the picture and whether the picture is cropped. Appendix Table 1 shows these results in more detail, with exact Likes and Ad Views displayed for each set of ads.

 The baseline photos on average have similar amounts of engagement, garnering almost identical levels of likes per view. This is important because if the baseline pictures were wildly different in terms of engagement, it would make interpretation of any results more challenging. (For example, if the baseline light complexion picture saw a 10\% engagement rate, and the dark complexion picture saw a 90\% engagement rate, then any differences from baseline would be hard to interpret). But when the ads zoom in, making skin more salient, the photos of people with lighter complexions receive a significantly higher boost relative to the photos of people with dark complexions. The boost is roughly twice as large: the engagement rates jump from 18.4\% to 30.8\% when the photos feature people with light complexions, as opposed to a jump of 18.3\% to 24.1\% for photos featuring people with dark complexions. Column 1 of Table 2 illustrates these results in a linear regression.

 This finding is robust across all six groups of the pictures tested. Appendix Table 2 shows the same results, disaggregated by photo. In the baseline (zoomed out, uncropped) images, we see no statistically significant difference in 3 of the 6 sets, a statistically significant bias towards darker complexions in 2 of the 6 sets, and a statistically significant bias towards the lighter complexion in the remaining 1 set. But in all six cases, when the ads zoom in, the picture with models with darker complexions is penalized more (or improves less), relative to its companion picture with models with lighter complexions.

\subsection{Is Meta Causing Disparities?}

The treatment effect could be explained by discriminatory users, by Meta’s decision about who sees the ads, or some combination of the two. As noted above, Meta is actively choosing to serve the ads to different populations who differ along observable demographics of gender and age (other demographics, such as race, are not shared with advertisers). This could help explain the disparities we find just as much as user behavior.

We can test this more accurately -- albeit imperfectly -- in two ways: by measuring the treatment effect after controlling for user demographics, and by assessing the treatment effect in an exploratory round of this experiment when Meta turned its audience optimization off.

First, we test whether the treatment effect persists after controlling for user demographics. If the treatment effect dissipates when controlling for observable demographics, this suggests it is Meta’s choice of audience that is driving the racial disparity. If the treatment effect is stable, then this suggests either that user behavior, not Meta audience decisions, are driving the disparity, or that Meta’s decisions are driving the disparity but in a way that we cannot observe.

We find that the disparity persists, even when controlling for audience characteristics. Table 2 presents the results of a regression measuring the treatment effect, when we do and do not control for observable demographics. Column 1 present a univariate analysis replicating the main treatment effect. Column 2 presents the same regression, but controlling for audience demographics. We find that a higher proportion of likes are independently associated with cropped image, whether the viewer is female, and the viewer’s age category. 


\begin{table}[ht]
\centering
\resizebox{\columnwidth}{!}{%
\begin{tabular}{@{}lcc@{}}
\toprule
                                         & \textbf{Model 1} & \textbf{Model 2} \\
\midrule
Intercept                                & 0.185  (0.004)   & 0.193  (0.010)   \\
Is Darker Complexion                     & -0.001  (0.006)  & -0.016 (0.014)   \\
Is Cropped                               & 0.122   (0.007)  & 0.098  (0.007)   \\
User is Female                           & --               & -0.032  (0.008)  \\
User Age Category (1 - 6)                & --               & 0.011 (0.003)    \\
Is Darker Complexion * Is Cropped        & -0.064 (0.009)   & -0.056  (0.010)  \\
Is Darker Complexion * User is Female    & --               & 0.001  (0.010)   \\
Is Darker Complexion * User Age Category & --               & 0.006 (0.004)    \\
Adj R2                                   & 0.013            & 0.017            \\
N                                        & 34,159           & 34,159           \\
\bottomrule
\end{tabular}%
}
\caption{Linear regression of engagement rate on photo and viewer characteristics. Standard errors clustered at the (picture complexion * cropped) level. The age variable is a categorical variable ranging from 1 to 6, with 1 being ages 18 - 24, 2 being ages 25-34, 3 being ages 35-44, 4 being ages 45-54, 5 being ages 55-64, and 6 being ages 65+.}
\end{table}

Table 2 shows that the main treatment effect holds, even when controlling for what we know about the audience demographics. Without controlling for audience demographics, we find a treatment effect of roughly 6.4 percentage points. When controlling for user gender and age category, the treatment effect diminishes slightly, to 5.6 percentage points, but is not statistically significantly different from the treatment effect when not controlling for audience demographics.

Second, we can see whether the treatment effect persists when Meta turns its audience optimization algorithms off. As described in the supplementary materials, we ran a version of this experiment across all 50 states to assess the geographic variation in the treatment effect. When we conducted this experiment, an unexpected outcome was that Meta sent a warning message that it would not be able to run its audience optimization algorithm because this advertising campaign was running so many simultaneous ads. Meta states that when an advertiser runs too many ads at once, 1,200 in our case, ads are unable to be optimized properly and can deliver less often with worse results. Meta recommended that for an account of our size, the upper limit for good performance is 250 ads.  

Because this was unexpected, we did not pre-register this test, so it should be considered exploratory. Nonetheless, we can see how ad engagement rates -- and the treatment effect -- changed during these tests. 

The average Like/Ad View is significantly lower in the geographic experiment (0.12) when compared to the main experiment (0.22), which suggests that Meta’s audience optimization tools are effective. However, the main treatment effect persists, with the light complexion pictures improving by 12\% when cropped, while the dark complexion pictures see a slight penalty of 0.8\% when cropped, as shown in Appendix Table 3.

There are important limitations to both approaches. The first test, which shows that the treatment effect persists when controlling for demographics, is limited because there are many user traits we do not observe, especially ethnicity, but also traits such as socioeconomic status and political attitudes. The second test, which shows that the treatment effect persists when Meta turns its audience optimization tool off and the ads are run in 50 states, is also limited because we have little information on what, exactly, Meta is doing when it sends this warning. Other optimization tools could still be on, for example. 

In sum, further testing helps tease out the role of user-level discrimination and Meta market design decisions on disparities. It provides provisional evidence that racial disparities are either driven by user preferences or by a combination of user preferences and Meta decisions that are not observable to the experimenters. 

\subsection{Estimating the Economic Cost of the Racial Penalty}
The difference in Like rates translates to economic penalties for advertisers who use models with dark complexions. Meta’s Ad Manager lets us calculate this cost more precisely. The basic question we assess is -- if one advertiser spends \$1000 promoting a photo of someone with a light complexion to get some level of engagement, how much more would she have to spend to get the same level of engagement if the photo highlighted a model with a dark complexion? 

Pictures of models with light complexions received Likes 3,880 times out of 16,784 Reaches (23.1\% of the time). The pictures of models with dark complexions received Likes 3,650 times out of 17,635 (20.7\% of the time). Hence, for every \$1,000 that an advertiser spends to achieve a fixed level of engagement for pictures with models with light complexions, an advertiser using pictures of models with dark complexions would have to spend 11.59\% more, or \$1,159, to achieve the same result.


There are limits in how to interpret this penalty. One could imagine a picture with models with darker skin that garners fewer Likes, but does better on some other metrics, like conversions. Perhaps fewer people click “Like”, but are more likely to then visit the wedding photographer’s page, send a message, and hire her. Hence, our calculation here needs to be interpreted with caution, as it only speaks to how much money an advertiser would have to spend to get a level of engagement defined in a specific way.

\subsection{Meta Budget Optimization Tool Leads to Racial Disparities in Ad Spending}

Advertisers can explicitly ask Meta for help in choosing how to spend their advertising budgets. The tool works as follows. Consider an advertiser who has two advertisements to show during a campaign. Suppose the advertiser does not know if one ad works better than the other. Meta helps the advertiser optimize her budget choices. First, both ads would be displayed to audiences, but Meta can learn based on user engagement if one ad is more effective and then funnel money towards that advertisement. We take advantage of this to test whether Meta’s optimization tool leads to different spending levels for photos that feature people with different skin complexions when the feature is ON versus when it is OFF. Importantly, this is an exploratory data analysis, since this test was not pre-registered and therefore not part of our main experiment.

When we do this, we find that Meta automatically funnels the advertising budget towards pictures of people with light skin complexions, presumably to maximize audience engagement (Appendix Figure 3, Appendix Tables 5-6). When budget optimization is turned off, Meta allocates the total budget identically across the four conditions (light complexion and uncropped, light complexion and cropped, dark complexion and uncropped, and dark complexion and cropped). Each condition receives roughly 25\% of the entire advertising budget. As a result, \$124.52 is spent on photos of models with light complexions, as compared to \$124.40 on photos of models with dark complexions. But when optimization is turned on, Meta automatically funnels money towards the photos of models with lighter skin complexions, which receive nearly two-thirds of the entire budget instead of half. \$159.43 is spent on photos of models with light complexions versus \$89.70 on photos of models with dark complexions. (While we do see photos of light skin models outperform dark skin models for both cropped and uncropped images, it is not clear why Meta funnels more money to uncropped images which receive lower likes/view.) Hence, disparities in outcomes driven by user preference can get amplified by Meta’s budget optimization tools. This finding is distinct from work by \cite{lambrecht2019algorithmic} which finds that even when an ad is designed to be shown in a gender-neutral way, cost optimization algorithms show the ad to more men because the male audience was less desirable (and therefore cheaper to target).

\subsection{Does Discrimination extend to Brazil?}

To assess the generalizability of our findings, we replicated the experiment in Brazil. We focus on Brazil for two reasons. First, it is important to extend audit studies beyond the United States and Western European countries. Second, while racism and colorism are severe problems in both countries, Brazil’s demographics are different from the United States’, so the mechanism underlying the results of our US-based experiment might manifest differently.

We use the same set of 24 ads featured in the main experiment, only changing the audience to Instagram users in Brazil, and translating the advertisement captions to Portuguese with the assistance of a native speaker. Other parameters such as ad budget, audience age, audience interest, and ad objective remain the same. The Brazilian audience parameters yielded an eligible audience in the range of 8,400,000 - 9,900,000. As this was an exploratory data analysis, the test was not pre-registered and was conducted independently of our main experiment.

In total, the 24 advertisements received 131,219 views and 18,963 likes, yielding an average Like/Ad View of .14. Demographic summary statistics for viewers are listed in Appendix Table 7. Similar to the main experiment, a majority of viewers are female. However, unlike in the main experiment, viewers aged 18-24 do not make up the majority of viewers.

Once again, our findings reveal a penalization effect for images featuring models with darker complexions. Regression results are displayed in Table 3, while exact Likes and Ad Views per experimental condition are presented in Appendix Tables 8 and 9 presents results segmented by photo.

Notably, in contrast to the main experiment, baseline photos did not exhibit comparable levels of engagement; uncropped ads featuring light complexions demonstrated a 1.9 percentage point higher Like-to-View ratio than their dark-complexion counterparts. Nevertheless, akin to our previous observations, when the ads zoomed in, accentuating skin, photos of individuals with lighter complexions experienced a significantly higher boost in engagement compared to those with darker complexions. On one hand, the lack of comparable baselines makes the Brazil results harder to interpret. On the other, it is nonetheless striking that the main finding holds. Taken together, the persistence of a bias towards light complexion photos suggests that our findings are robust across large geographical regions. 


\begin{table}[ht]
\centering
\resizebox{\columnwidth}{!}{%
\begin{tabular}{@{}lcc@{}}
\toprule
                                         & \textbf{Model 1} & \textbf{Model 2} \\
\midrule
Intercept                                & 0.141 (0.008)    & 0.117 (0.008)    \\
Is Darker Complexion                     & -0.019 (0.011)   & -0.029 (0.011)   \\
Is Cropped                               & 0.041 (0.000)    & 0.035 (0.000)    \\
User is Female                           & --               & -0.023 (0.002)   \\
User Age Category (1 - 6)                & --               & 0.014 (0.002)    \\
Is Darker Complexion * Is Cropped        & -0.013 (0.000)   & -0.015 (0.000)   \\
Is Darker Complexion * User is Female    & --               & -0.002 (0.002)   \\
Is Darker Complexion * User Age Category & --               & 0.006 (0.003)    \\
Adj R2                                   & 0.003            & 0.009            \\
N                                        & 130,920          & 130,920          \\
\bottomrule
\end{tabular}%
}
\caption{Linear regression of engagement rate on photo and viewer characteristics for Brazil experiment. Standard errors clustered at the (picture complexion * cropped) level. The age variable is a categorical variable ranging from 1 to 6, with 1 being ages 18 - 24, 2 being ages 25-34, 3 being ages 35-44, 4 being ages 45-54, 5 being ages 55-64, and 6 being ages 65+.}
\end{table}

\section{Discussion}

This paper presents the results of an experiment measuring racial bias in a dominant online advertising market. Given two identical or nearly-identical photographs that vary by the model’s skin complexion, making the skin more salient by cropping the picture leads to a 21.75\% penalty when the model’s skin is darker. This directly translates to higher costs for advertisers who feature people with dark skin complexions. We estimate that the racial penalty is associated with a rise in advertising costs of 11.59\%.

One contribution from this paper is that it finds a racial disparity along a dimension – likes – that is likely to be more related to animus than to other factors \cite{becker1957economics}. In many audit studies, racial penalties are found in contexts where the person discriminating might well be making a (perhaps errant, perhaps odious) statistical inference using race as a proxy for some other outcome \cite{bohren2019dynamics}. Here, the discrimination we document is simple and instantaneous -- tapping a “Like” button in response to a photograph. If tapping “Like” signifies nothing more than an aesthetic response to a picture, then this is a clean measure of taste-based discrimination. If users tap “Like” for other reasons -- such as to manipulate the types of advertisements that Meta shows -- then the disparities we find would be driven by something more complicated.

This paper also contributes to the literature seeking a novel method to measure racial animus at scale. This is an increasing challenge because racism is, more and more, too unacceptable to admit to publicly. Across geographies, our measure of a racial penalty can be measured at scale and shows little correlation with existing measures, such as self-reported racial attitudes. 

Finally, the experiment also highlights new challenges that regulators and marketplace designers face in addressing a very old problem. Has Meta done anything illegal? The answer is far from clear. In one sense, Meta is a market-place designer offering an effective way to reach audiences. Meta plays a large role in how to serve these ads, trying to optimize for engagement and helping advertisers optimize their budgets. We find strong evidence that Meta manipulates which audiences see the ad, and that this choice of audience differs based on skin complexion. But we do not find clear evidence that this improves or worsens engagement rates by skin complexion. We find stronger evidence that Meta’s budget optimization tool does, indeed, exacerbate discrimination, by funneling advertiser money towards photos of models with lighter complexions.

Limits to the experiment suggest further avenues of research. Past research has documented how colorism and racism both operate to harm African Americans in the criminal justice system \cite{wickett2021not}. This experiment does not disentangle the two. It most directly tests the impact of colorism by manipulating skin tone, but because skin complexion is a proxy for racial groups in the United States, our findings could also be driven by racism, either against African Americans or other groups with darker skin tones. This is a common limitation in audit studies, which always focus on manipulating a feature like name or physical appearance in a way that proxies for race. Future research could explore this, even with the same experimental design. In addition, as with other audit studies, this experiment captures a harm whose longer-term implications are unclear. If African-American applicants don’t get callbacks for job interviews, does this translate to lower wages? If models with darker skin complexions are penalized in the engagement with their ads, does this lead to less work? Understanding these dynamics is important, but outside the scope of this paper.

In sum, this paper explores how an old problem evolves when its market environment changes. User-level discriminatory attitudes are nothing new, nor is targeted advertising. What is new is the way that facially neutral algorithmic decision-making can mix with user-level discrimination to pose new legal and ethical challenges.

\bibliography{aaai25}

\clearpage 
\appendix 

\section*{Appendix}

\subsection{Geographic Variation in Discrimination Measure}
We repeat our main experiment, placing each of the 24 ads in the 50 states with an equal budget of \$19 across states to assess state-level racial attitudes.

In total, 43,885 Instagram users viewed one of the 1,200 advertisements and sent 5,317 likes. Each state recorded an average of 868 advertisement views and 106 likes for an average like/ad view of 0.12 (Figure S1, Table S3-S4). The 5 states with the lowest $(P_{LZ}-P_{DZ})-(P_L-P_D)$ values -- indicating lower animus towards darker skin complexions -- are Wisconsin, South Carolina, Tennessee, Minnesota, and Utah. The 5 states with the highest (PLZ-PDZ)-(PL-PD) value are North Carolina, Massachusetts, Florida, Georgia, and Pennsylvania.

We then compare how our metric varies across states to the state-level variance of seven other metrics of race animus. Four are survey based metrics: the Project Implicit self-survey of Black-White racial attitudes, a measurement of racial resentment derived from the American National Election Studies survey (MrP), and two metrics derived from the Cooperative Congressional Election Study (Racial Resentment and Racial Resentment Among Whites). One is a non-survey internet based measure: the popularity of racially charged language on Google search. Another is the number of hate groups in a state. And lastly, we use data on email response rates to Black and White senders from a recent large scale field experiment by \cite{block2021americans}.

\cite{block2021americans} find that most measures of racial animus are not correlated with each other; each measure may be highlighting a specific independent mechanism of discrimination. Similarly, we find our new metric is not highly correlated with any of these existing measures. These results are shown in Figure S2. All data for these metrics come from the replication data of \cite{block2021americans}.

\begin{table}[H]
\renewcommand{\tablename}{Appendix Table}
\resizebox{\columnwidth}{!}{%
\begin{tabular}{|l|l|l|l|}
\hline
                   & \multicolumn{1}{c|}{\textbf{Dark Complexion}}                  & \multicolumn{1}{c|}{\textbf{Light Complexion}}                 & \multicolumn{1}{c|}{\textbf{Total}}                            \\ \hline
\textbf{Cropped}   & \begin{tabular}[c]{@{}l@{}}1735/7189\\ (24.1\%)\end{tabular}   & \begin{tabular}[c]{@{}l@{}}1954/6335\\ (30.8\%)\end{tabular}   & \begin{tabular}[c]{@{}l@{}}3689/13524 \\ (27.3\%)\end{tabular} \\ \hline
\textbf{Uncropped} & \begin{tabular}[c]{@{}l@{}}1915/10446\\ (18.3\%)\end{tabular}  & \begin{tabular}[c]{@{}l@{}}1926/10449\\ (18.4\%)\end{tabular}  & \begin{tabular}[c]{@{}l@{}}3841/20895 \\ (18.4\%)\end{tabular} \\ \hline
\textbf{Total}     & \begin{tabular}[c]{@{}l@{}}3650/17635 \\ (20.7\%)\end{tabular} & \begin{tabular}[c]{@{}l@{}}3880/16784 \\ (23.1\%)\end{tabular} & \begin{tabular}[c]{@{}l@{}}7530/34419 \\ (21.9\%)\end{tabular} \\ \hline
\end{tabular}%
}
\caption{Likes per Ad Views by Experimental Condition. This figure shows the engagement rate for each experimental group. Specifically, of all advertisement Views, how many times did a viewer “Like” the picture. All comparisons were significant at p<.001 except for Dark Complexion uncropped vs Light Complexion uncropped (p=0.88).}
\end{table}

\begin{table}[]
\renewcommand{\tablename}{Appendix Table}
\renewcommand{\arraystretch}{1.2}
\resizebox{\columnwidth}{!}{%
\begin{tabular}{|l|l|l|l|l|l|}
\hline
\multicolumn{1}{|c|}{\textbf{Groups}} & \multicolumn{1}{c|}{\textbf{Image}} & \multicolumn{1}{c|}{\textbf{Likes / Views}} & \multicolumn{1}{c|}{\textbf{Image}} & \multicolumn{1}{c|}{\textbf{Likes / Views}} & \multicolumn{1}{c|}{\textbf{P-value}} \\ \hline
1 & \begin{tabular}[c]{@{}l@{}}Dark Complexion torso, \\ uncropped\end{tabular} & \begin{tabular}[c]{@{}l@{}}275/1276 \\ (21.6\%)\end{tabular} & \begin{tabular}[c]{@{}l@{}}Light Complexion torso, \\ uncropped\end{tabular} & \begin{tabular}[c]{@{}l@{}}240/1479 \\ (16.2\%)\end{tabular} & \textless{}.001 \\ \hline
1 & \begin{tabular}[c]{@{}l@{}}Dark Complexion torso, \\ cropped\end{tabular} & \begin{tabular}[c]{@{}l@{}}120/539\\ (22.3\%)\end{tabular} & \begin{tabular}[c]{@{}l@{}}Light Complexion torso, \\ cropped\end{tabular} & \begin{tabular}[c]{@{}l@{}}139/474  \\ (29.3\%)\end{tabular} & 0.01 \\ \hline
2 & \begin{tabular}[c]{@{}l@{}}Dark Complexion holding flowers, \\ uncropped\end{tabular} & \begin{tabular}[c]{@{}l@{}}258/1107\\ (23.3\%)\end{tabular} & \begin{tabular}[c]{@{}l@{}}Light Complexion holding flowers, \\ uncropped\end{tabular} & \begin{tabular}[c]{@{}l@{}}271/1162 \\ (23.3\%)\end{tabular} & 0.993 \\ \hline
2 & \begin{tabular}[c]{@{}l@{}}Dark Complexion holding flowers, \\ cropped\end{tabular} & \begin{tabular}[c]{@{}l@{}}178/852 \\ (20.9\%)\end{tabular} & \begin{tabular}[c]{@{}l@{}}Light Complexion holding flowers, \\ cropped\end{tabular} & \begin{tabular}[c]{@{}l@{}}209/674\\ (31\%)\end{tabular} & \textless{}.001 \\ \hline
3 & \begin{tabular}[c]{@{}l@{}}Dark Complexion holding hands, \\ uncropped\end{tabular} & \begin{tabular}[c]{@{}l@{}}153/483  \\ (31.7\%)\end{tabular} & \begin{tabular}[c]{@{}l@{}}Light Complexion holding hands, \\ uncropped\end{tabular} & \begin{tabular}[c]{@{}l@{}}173/709 \\ (24.4\%)\end{tabular} & 0.006 \\ \hline
3 & \begin{tabular}[c]{@{}l@{}}Dark Complexion holding hands, \\ cropped\end{tabular} & \begin{tabular}[c]{@{}l@{}}178/938\\ (19.0\%)\end{tabular} & \begin{tabular}[c]{@{}l@{}}Light Complexion holding hands, \\ cropped\end{tabular} & \begin{tabular}[c]{@{}l@{}}159/696\\ (22.8\%)\end{tabular} & 0.056 \\ \hline
4 & \begin{tabular}[c]{@{}l@{}}Dark Complexion keyhole dress, \\ uncropped\end{tabular} & \begin{tabular}[c]{@{}l@{}}450/2665  \\ (16.9\%)\end{tabular} & \begin{tabular}[c]{@{}l@{}}Light Complexion keyhole dress, \\ uncropped\end{tabular} & \begin{tabular}[c]{@{}l@{}}490/2711\\ (18.1\%)\end{tabular} & 0.251 \\ \hline
4 & \begin{tabular}[c]{@{}l@{}}Dark Complexion keyhole dress, \\ cropped\end{tabular} & \begin{tabular}[c]{@{}l@{}}312/920  \\ (33.9\%)\end{tabular} & \begin{tabular}[c]{@{}l@{}}Light Complexion keyhole dress, \\ cropped\end{tabular} & \begin{tabular}[c]{@{}l@{}}300/711 \\ (42.2\%)\end{tabular} & \textless{}.001 \\ \hline
5 & \begin{tabular}[c]{@{}l@{}}Dark Complexion racerback dress, \\ uncropped\end{tabular} & \begin{tabular}[c]{@{}l@{}}416/2445 \\ (17\%)\end{tabular} & \begin{tabular}[c]{@{}l@{}}Light Complexion racerback dress, \\ uncropped\end{tabular} & \begin{tabular}[c]{@{}l@{}}427/2772 \\ (15.4\%)\end{tabular} & 0.115 \\ \hline
5 & \begin{tabular}[c]{@{}l@{}}Dark Complexion racerback dress, \\ cropped\end{tabular} & \begin{tabular}[c]{@{}l@{}}652/2782 \\ (23.4\%)\end{tabular} & \begin{tabular}[c]{@{}l@{}}Light Complexion racerback dress, \\ cropped\end{tabular} & \begin{tabular}[c]{@{}l@{}}834/2887  \\ (28.9\%)\end{tabular} & \textless{}.001 \\ \hline
6 & \begin{tabular}[c]{@{}l@{}}Dark Complexion V-back dress, \\ uncropped\end{tabular} & \begin{tabular}[c]{@{}l@{}}363/2470\\ (14.7\%)\end{tabular} & \begin{tabular}[c]{@{}l@{}}Light Complexion V-back dress, \\ uncropped\end{tabular} & \begin{tabular}[c]{@{}l@{}}325/1616  \\ (20.1\%)\end{tabular} & \textless{}.001 \\ \hline
6 & \begin{tabular}[c]{@{}l@{}}Dark Complexion V-back dress, \\ cropped\end{tabular} & \begin{tabular}[c]{@{}l@{}}295/1158  \\ (25.5\%)\end{tabular} & \begin{tabular}[c]{@{}l@{}}Light Complexion V-back dress, \\ cropped\end{tabular} & \begin{tabular}[c]{@{}l@{}}313/893 \\ (35.1\%)\end{tabular} & \textless{}.001 \\ \hline
\end{tabular}%
}
\caption{Results Disaggregated by Photo. This Table shows the engagement rate for each experimental advertisement. Specifically, of all advertisement Views, how many times did a viewer “Like” the picture. In the baseline (zoomed out, uncropped) images, we see no statistically significant difference in 3 of the 6 sets, a statistically significant bias towards darker complexions in 2 of the 6 sets, and a statistically significant bias towards the lighter complexion in the remaining 1 set. But in all six cases, when the ads zoom in, the picture with models with darker complexions is penalized more (or improves less), relative to its companion picture with models with lighter complexions.}
\end{table}

\begin{figure*}[]
\renewcommand{\figurename}{Appendix Figure}
\centering
\includegraphics[width=0.5\textwidth]{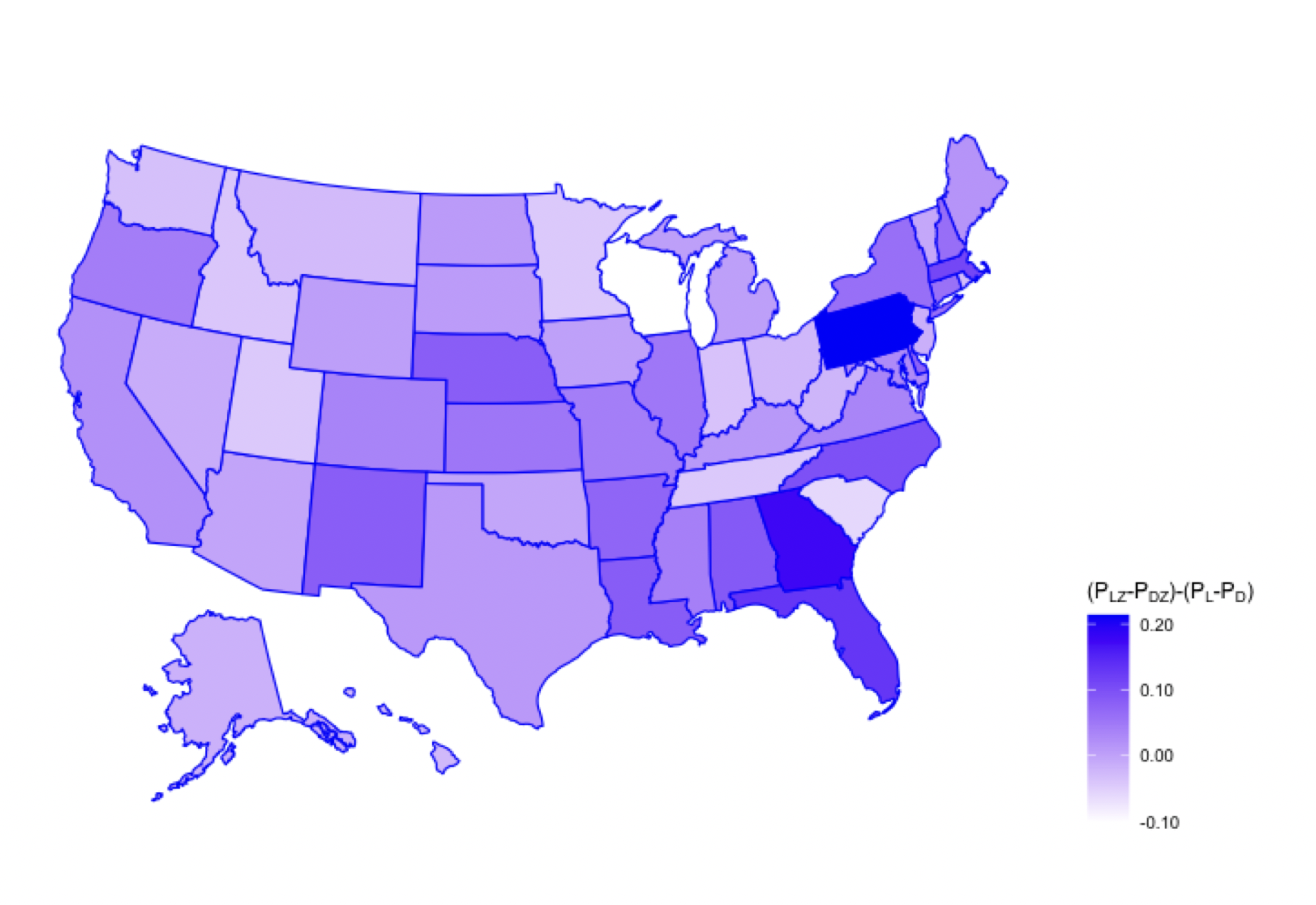} 
\caption{$(P_{LZ}-P_{DZ})-(P_L-P_D)$ by State. This Figure displays the measure of racial attitudes drawn from our treatment effect by viewer’s location at the state level. Darker areas correspond to areas with higher animus towards darker skin complexion advertisements.}
\label{figs1}
\end{figure*}

\begin{figure*}[]
\renewcommand{\figurename}{Appendix Figure}
\centering
\includegraphics[width=0.5\textwidth]{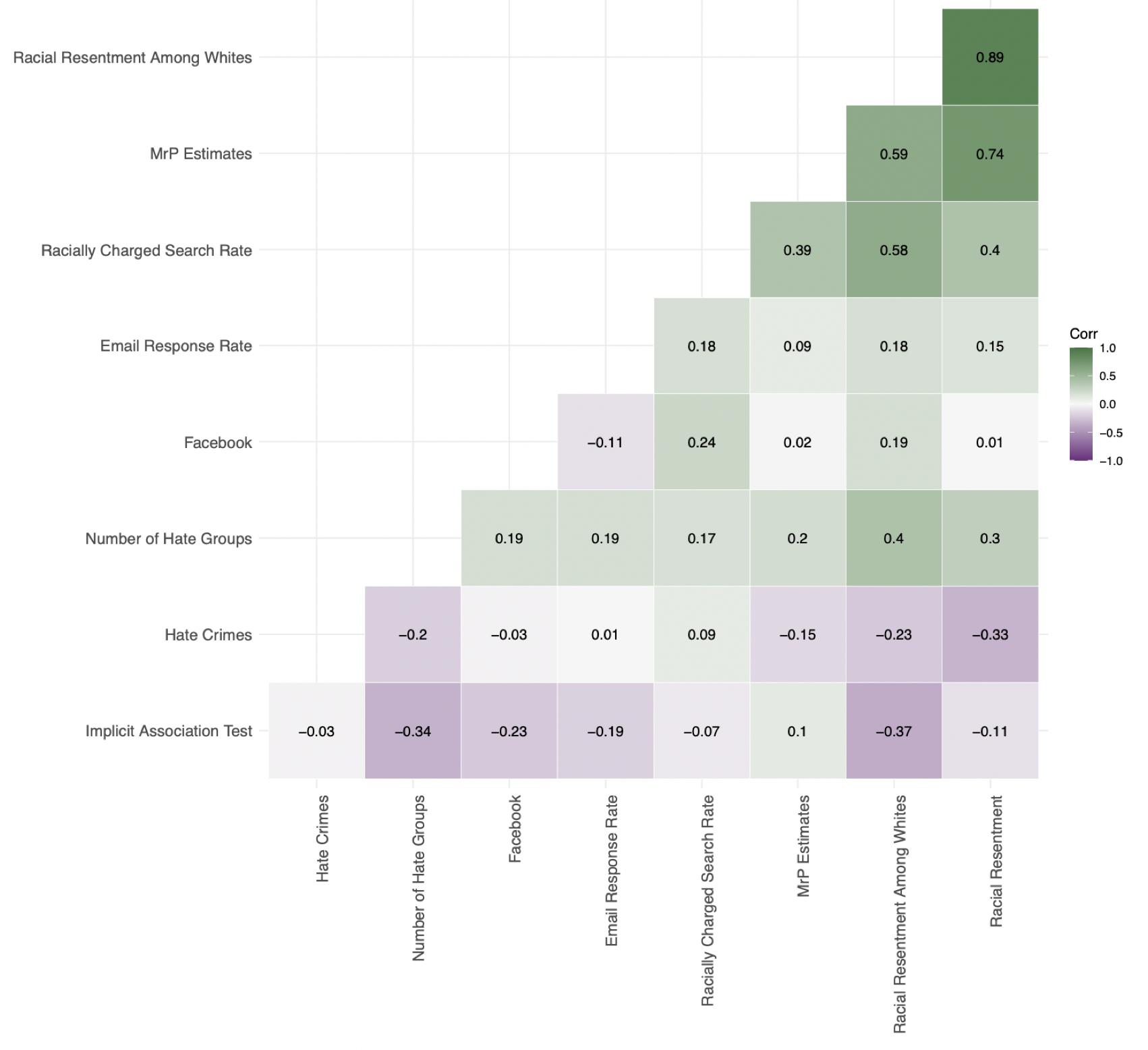} 
\caption{Measures of Racial Animus by State. This Figure displays various measures of racial attitudes including our treatment effect by viewer’s location at the state level. Purple cells correspond to negative correlations and green cells correspond to positive correlations. Block et al. highlight that apart from metrics that are functions of one another such as MrP and Racial Resentment, most measures of racial animus are not strongly correlated.}
\label{figs2}
\end{figure*}

\begin{table}[]
\renewcommand{\tablename}{Appendix Table}
\resizebox{\columnwidth}{!}{%
\begin{tabular}{|l|l|l|l|}
\hline
\textbf{} & \textbf{Dark Complexion} & \textbf{Light Complexion} & \textbf{Total} \\ \hline
\textbf{Cropped} & \begin{tabular}[c]{@{}l@{}}1304/11488\\ (11.4\%)\end{tabular} & \begin{tabular}[c]{@{}l@{}}1094/7783\\(14.1\%)\end{tabular} & \begin{tabular}[c]{@{}l@{}}2398/19271 \\(12.4\%)\end{tabular} \\ \hline
\textbf{Uncropped} & \begin{tabular}[c]{@{}l@{}}1321/11439\\(11.5\%)\end{tabular} & \begin{tabular}[c]{@{}l@{}}1598/12675\\(12.6\%)\end{tabular} & \begin{tabular}[c]{@{}l@{}}2919/24114\\(12.1\%)\end{tabular} \\ \hline
\textbf{Total} & \begin{tabular}[c]{@{}l@{}}2625/22927 \\(11.4\%)\end{tabular} & \begin{tabular}[c]{@{}l@{}}2692/20458 \\(13.2\%)\end{tabular} & \begin{tabular}[c]{@{}l@{}}5317/43385\\  (12.1\%)\end{tabular} \\ \hline
\end{tabular}%
}
\caption{Aggregated Likes / Ad Views: This table shows the treatment effect for a round of the experiment where the 24 ads were separately run across all 50 states to measure geographic variation. During this experiment, Meta warned that its audience optimization tools would be turned off because of the number of ads run simultaneously. Compared to the main experimental round, we find lower audience engagement rates, but we still find a treatment effect. All comparisons across complexions were significant at p<.001 except for Dark Complexion uncropped vs Light Complexion uncropped (p=.012).}
\end{table}

\begin{table}[]
\renewcommand{\tablename}{Appendix Table}
\resizebox{\columnwidth}{!}{%
\begin{tabular}{|l|l|l|l|}
\hline
\textbf{State} & \textbf{$(P_{LZ}-P_{DZ})-(P_L-P_D)$} & \textbf{Likes} & \textbf{Reach} \\ \hline
United States & 0.016 & 5317 & 43385 \\ \hline
Alabama & 0.088 & 111 & 905 \\ \hline
Alaska & -0.019 & 35 & 1020 \\ \hline
Arizona & -0.005 & 134 & 876 \\ \hline
Arkansas & 0.072 & 93 & 869 \\ \hline
California & 0.02 & 205 & 890 \\ \hline
Colorado & 0.035 & 120 & 735 \\ \hline
Connecticut & 0.057 & 124 & 890 \\ \hline
Delaware & 0.075 & 57 & 888 \\ \hline
Florida & 0.131 & 189 & 748 \\ \hline
Georgia & 0.176 & 145 & 783 \\ \hline
Hawaii & -0.028 & 85 & 1217 \\ \hline
Idaho & -0.042 & 71 & 970 \\ \hline
Illinois & 0.05 & 156 & 795 \\ \hline
Indiana & -0.039 & 104 & 856 \\ \hline
Iowa & 0 & 78 & 864 \\ \hline
Kansas & 0.051 & 82 & 869 \\ \hline
Kentucky & 0.01 & 100 & 884 \\ \hline
Louisiana & 0.083 & 115 & 939 \\ \hline
Maine & 0.015 & 72 & 874 \\ \hline
Maryland & 0.027 & 141 & 759 \\ \hline
Massachusetts & 0.111 & 124 & 759 \\ \hline
Michigan & 0.002 & 147 & 729 \\ \hline
Minnesota & -0.046 & 93 & 804 \\ \hline
Mississippi & 0.04 & 106 & 949 \\ \hline
Missouri & 0.042 & 126 & 881 \\ \hline
Montana & -0.029 & 53 & 948 \\ \hline
Nebraska & 0.082 & 76 & 995 \\ \hline
Nevada & -0.014 & 98 & 901 \\ \hline
New Hampshire & 0.061 & 72 & 883 \\ \hline
New Jersey & -0.017 & 174 & 727 \\ \hline
New Mexico & 0.082 & 74 & 754 \\ \hline
New York & 0.059 & 204 & 783 \\ \hline
North Carolina & 0.098 & 155 & 868 \\ \hline
North Dakota & 0.007 & 26 & 1015 \\ \hline
Ohio & -0.026 & 117 & 755 \\ \hline
Oklahoma & -0.006 & 93 & 884 \\ \hline
Oregon & 0.043 & 92 & 707 \\ \hline
Pennsylvania & 0.214 & 155 & 767 \\ \hline
Rhode Island & -0.03 & 71 & 864 \\ \hline
South Carolina & -0.062 & 119 & 983 \\ \hline
South Dakota & 0.005 & 39 & 1123 \\ \hline
Tennessee & -0.046 & 147 & 956 \\ \hline
Texas & 0.01 & 179 & 751 \\ \hline
Utah & -0.045 & 93 & 741 \\ \hline
Vermont & -0.008 & 38 & 897 \\ \hline
Virginia & 0.033 & 128 & 796 \\ \hline
Washington & -0.036 & 101 & 876 \\ \hline
West Virginia & -0.023 & 82 & 894 \\ \hline
Wisconsin & -0.103 & 86 & 802 \\ \hline
Wyoming & 0.002 & 32 & 962 \\ \hline
\end{tabular}%
}
\caption{Discrimination By State. This table shows our measure of discrimination by the viewer’s location at the state level, as measured by Meta. The measure matches our definition of the treatment effect in the experiment: it looks at how much any baseline racial disparities between zoomed-out pictures worsen.}
\end{table}

\begin{figure*}[]
\renewcommand{\figurename}{Appendix Figure}
\centering
\includegraphics[width=0.8\textwidth]{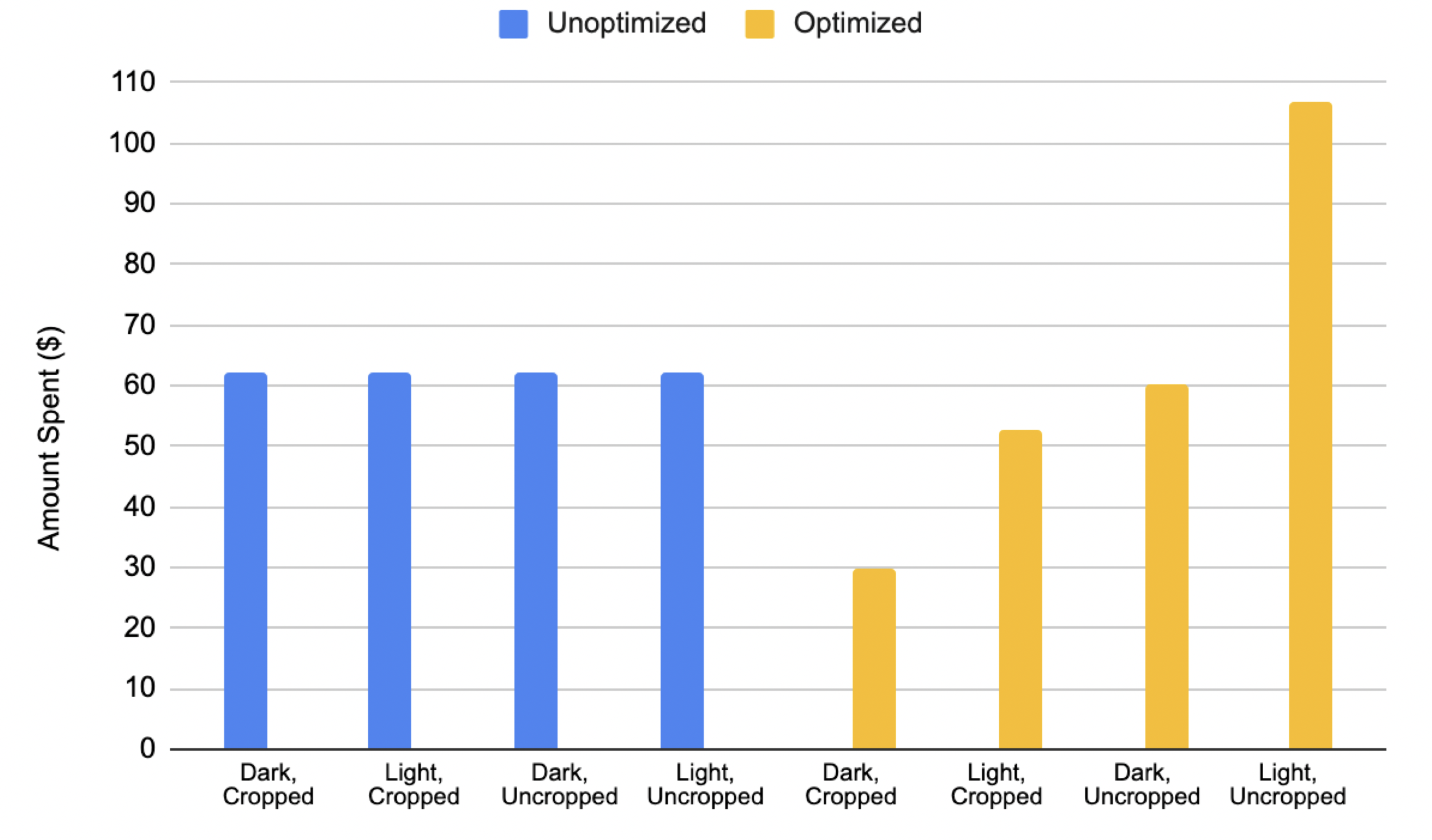} 
\caption{Optimization Results: This Figure measures how advertising budget is distributed across different pictures when Meta’s budget optimization feature is turned on or off. P-value < 0.001 for Chi-Squared test comparing Amount Spent in first four bars (Optimization OFF) versus Amount Spent in last four bars (Optimization ON).}
\label{figs3}
\end{figure*}

\begin{table}[]
\renewcommand{\tablename}{Appendix Table}
\renewcommand{\arraystretch}{1.5}
\resizebox{\columnwidth}{!}{%
\begin{tabular}{|l|l|l|r|r|r|r|}
\hline
\multicolumn{1}{|c|}{\textbf{Optimization}} & \multicolumn{1}{c|}{\textbf{Complexion}} & \multicolumn{1}{c|}{\textbf{Cropping}} & \multicolumn{1}{c|}{\textbf{Likes}} & \multicolumn{1}{c|}{\textbf{Reach}} & \multicolumn{1}{c|}{\textbf{Likes/Reach (\%)}} & \multicolumn{1}{c|}{\textbf{Amount Spent (\$)}} \\ \hline
Optimized & Dark & Cropped & 359 & 1529 & 23.48 & 29.67 \\ \hline
Optimized & Dark & Uncropped & 620 & 2917 & 21.25 & 60.03 \\ \hline
Optimized & Light & Cropped & 972 & 3488 & 27.87 & 52.67 \\ \hline
Optimized & Light & Uncropped & 1357 & 6322 & 21.46 & 106.76 \\ \hline
Unoptimized & Dark & Cropped & 583 & 1970 & 29.59 & 62.19 \\ \hline
Unoptimized & Dark & Uncropped & 644 & 2766 & 23.28 & 62.21 \\ \hline
Unoptimized & Light & Cropped & 612 & 1734 & 35.29 & 62.27 \\ \hline
Unoptimized & Light & Uncropped & 661 & 2642 & 25.02 & 62.25 \\ \hline
\end{tabular}%
}
\caption{Optimization Results. This Table measures how advertising budget is distributed across different picture classes when Meta’s budget optimization feature is turned on or off. P-value < 0.001 for Chi-Squared test comparing Amount Spent in first four rows (Optimization OFF) versus Amount Spent in last four rows (Optimization ON).}
\end{table}

\begin{table}[]
\renewcommand{\tablename}{Appendix Table}
\resizebox{\columnwidth}{!}{%
\begin{tabular}{|l|l|l|l|l|r|r|}
\hline
\textbf{Ad Name} & \textbf{Optimization} & \textbf{Complexion} & \textbf{Cropping} & \textbf{Likes/Reach (\%)} & \multicolumn{1}{l|}{\textbf{Amount Spent (\$)}} & \multicolumn{1}{l|}{\textbf{\begin{tabular}[c]{@{}l@{}}\% of dollars \\  spent in group\end{tabular}}} \\ \hline
flowers & optimized & dark & cropped & 9/28 (32.1\%) & 0.99 & 2.4 \\ \hline
flowers & optimized & dark & uncropped & 70/273 (25.6\%) & 6.16 & 14.8 \\ \hline
flowers & optimized & light & cropped & 66/243 (27.2\%) & 6.98 & 16.8 \\ \hline
flowers & optimized & light & uncropped & 359/1368 (26.2\%) & 27.44 & 66 \\ \hline
flowers & unoptimized & dark & cropped & 84/223 (37.7\%) & 10.37 & 25 \\ \hline
flowers & unoptimized & dark & uncropped & 109/458 (23.8\%) & 10.38 & 25 \\ \hline
flowers & unoptimized & light & cropped & 83/218 (38.1\%) & 10.38 & 25 \\ \hline
flowers & unoptimized & light & uncropped & 133/462 (28.8\%) & 10.39 & 25 \\ \hline
hands & optimized & dark & uncropped & 11/35 (31.4\%) & 1.7 & 4.1 \\ \hline
hands & optimized & dark & cropped & 152/559 (27.2\%) & 19.31 & 46.5 \\ \hline
hands & optimized & light & uncropped & 68/140 (48.6\%) & 8.3 & 20 \\ \hline
hands & optimized & light & cropped & 96/329 (29.2\%) & 12.18 & 29.4 \\ \hline
hands & unoptimized & dark & uncropped & 63/153 (41.2\%) & 10.39 & 25.1 \\ \hline
hands & unoptimized & dark & cropped & 88/255 (34.5\%) & 10.31 & 24.9 \\ \hline
hands & unoptimized & light & uncropped & 74/150 (49.3\%) & 10.34 & 25 \\ \hline
hands & unoptimized & light & cropped & 84/185 (45.4\%) & 10.4 & 25.1 \\ \hline
keyhole & optimized & dark & uncropped & 184/937 (19.6\%) & 14.27 & 34.3 \\ \hline
keyhole & optimized & dark & cropped & 0/3 (0\%) & 0.02 & 0 \\ \hline
keyhole & optimized & light & cropped & 1/2 (50\%) & 0.14 & 0.3 \\ \hline
keyhole & optimized & light & uncropped & 459/2639 (17.4\%) & 27.15 & 65.3 \\ \hline
keyhole & unoptimized & dark & cropped & 71/157 (45.2\%) & 10.4 & 25 \\ \hline
keyhole & unoptimized & dark & uncropped & 126/552 (22.8\%) & 10.39 & 25 \\ \hline
keyhole & unoptimized & light & cropped & 71/201 (35.3\%) & 10.35 & 24.9 \\ \hline
keyhole & unoptimized & light & uncropped & 128/642 (19.9\%) & 10.4 & 25 \\ \hline
racerback & optimized & dark & uncropped & 1/1 (100\%) & 0 & 0 \\ \hline
racerback & optimized & dark & cropped & 198/921 (21.5\%) & 9.2 & 22.2 \\ \hline
racerback & optimized & light & cropped & 795/2888 (27.5\%) & 32.21 & 77.7 \\ \hline
racerback & optimized & light & uncropped & 1/7 (14.3\%) & 0.05 & 0.1 \\ \hline
racerback & unoptimized & dark & cropped & 218/889 (24.5\%) & 10.38 & 25 \\ \hline
racerback & unoptimized & dark & uncropped & 141/606 (23.3\%) & 10.41 & 25.1 \\ \hline
racerback & unoptimized & light & cropped & 238/845 (28.2\%) & 10.38 & 25 \\ \hline
racerback & unoptimized & light & uncropped & 135/649 (20.8\%) & 10.38 & 25 \\ \hline
torso & optimized & dark & uncropped & 65/390 (16.7\%) & 7.3 & 17.5 \\ \hline
torso & optimized & dark & cropped & 0/0 (0\%) & 0 & 0 \\ \hline
torso & optimized & light & uncropped & 374/1784 (21\%) & 34.3 & 82.5 \\ \hline
torso & optimized & light & cropped & 0/0 (0\%) & 0 & 0 \\ \hline
torso & unoptimized & dark & cropped & 49/100 (49\%) & 10.32 & 24.9 \\ \hline
torso & unoptimized & dark & uncropped & 107/558 (19.2\%) & 10.35 & 25 \\ \hline
torso & unoptimized & light & cropped & 63/110 (57.3\%) & 10.41 & 25.1 \\ \hline
torso & unoptimized & light & uncropped & 104/491 (21.2\%) & 10.34 & 25 \\ \hline
v-back & optimized & dark & uncropped & 289/1281 (22.6\%) & 30.6 & 73.9 \\ \hline
v-back & optimized & dark & cropped & 0/18 (0\%) & 0.15 & 0.4 \\ \hline
v-back & optimized & light & cropped & 14/26 (53.8\%) & 1.16 & 2.8 \\ \hline
v-back & optimized & light & uncropped & 96/384 (25\%) & 9.52 & 23 \\ \hline
v-back & unoptimized & dark & uncropped & 98/439 (22.3\%) & 10.29 & 24.8 \\ \hline
v-back & unoptimized & dark & cropped & 73/346 (21.1\%) & 10.41 & 25.1 \\ \hline
v-back & unoptimized & light & cropped & 73/175 (41.7\%) & 10.35 & 25 \\ \hline
v-back & unoptimized & light & uncropped & 87/248 (35.1\%) & 10.4 & 25.1 \\ \hline
\end{tabular}%
}
\caption{Optimization Results Disaggregated by Photo. This Table measures how advertising budget is distributed across different pictures when Meta’s budget optimization feature is turned on or off.}
\end{table}

\begin{table}[]
\renewcommand{\tablename}{Appendix Table}
\resizebox{\columnwidth}{!}{%
\begin{tabular}{|r|l|llll|}
\hline
\multicolumn{1}{|l|}{\textbf{}} & \textbf{} & \multicolumn{4}{c|}{\textbf{Ad Views}} \\ \hline
\multicolumn{1}{|l|}{\textbf{}} & \textbf{} & \multicolumn{2}{c|}{\textbf{Dark Complexion}} & \multicolumn{2}{c|}{\textbf{Light Complexion}} \\ \hline
\multicolumn{1}{|l|}{\textbf{}} & \textbf{All Ads} & \multicolumn{1}{l|}{\textbf{Cropped}} & \multicolumn{1}{l|}{\textbf{Uncropped}} & \multicolumn{1}{l|}{\textbf{Cropped}} & \textbf{Uncropped} \\ \hline
\multicolumn{1}{|l|}{\textbf{Gender}} &  & \multicolumn{1}{l|}{} & \multicolumn{1}{l|}{} & \multicolumn{1}{l|}{} &  \\ \hline
Female & 98,596 (75\%) & \multicolumn{1}{l|}{19,919 (65\%)} & \multicolumn{1}{l|}{33,098 (81\%)} & \multicolumn{1}{l|}{15,652 (64\%)} & 29,927 (86\%) \\ \hline
Male & 32,324 (25\%) & \multicolumn{1}{l|}{10,883 (35\%)} & \multicolumn{1}{l|}{7,625 (19\%)} & \multicolumn{1}{l|}{8,863 (36\%)} & 4,953 (14\%) \\ \hline
Unknown & 299 (0.2\%) & \multicolumn{1}{l|}{71 (0.2\%)} & \multicolumn{1}{l|}{114 (0.3\%)} & \multicolumn{1}{l|}{52 (0.2\%)} & 62 (0.2\%) \\ \hline
\multicolumn{1}{|l|}{\textbf{Age}} &  & \multicolumn{1}{l|}{} & \multicolumn{1}{l|}{} & \multicolumn{1}{l|}{} &  \\ \hline
18-24 & 24,897 (19\%) & \multicolumn{1}{l|}{5,530 (18\%)} & \multicolumn{1}{l|}{10,210 (25\%)} & \multicolumn{1}{l|}{3,461 (14\%)} & 5,696 (16\%) \\ \hline
25-34 & 33,158 (25\%) & \multicolumn{1}{l|}{8,169 (26\%)} & \multicolumn{1}{l|}{10,693 (26\%)} & \multicolumn{1}{l|}{6,015 (24\%)} & 8,281 (24\%) \\ \hline
35-44 & 30,119 (23\%) & \multicolumn{1}{l|}{7,561 (24\%)} & \multicolumn{1}{l|}{8,451 (21\%)} & \multicolumn{1}{l|}{5,894 (24\%)} & 8,213 (24\%) \\ \hline
45-54 & 23,260 (18\%) & \multicolumn{1}{l|}{5,184 (17\%)} & \multicolumn{1}{l|}{6,407 (16\%)} & \multicolumn{1}{l|}{4,728 (19\%)} & 6,941 (20\%) \\ \hline
55-64 & 12,978 (9.9\%) & \multicolumn{1}{l|}{2,916 (9.4\%)} & \multicolumn{1}{l|}{3,383 (8.3\%)} & \multicolumn{1}{l|}{2,873 (12\%)} & 3,806 (11\%) \\ \hline
65+ & 6,807 (5.2\%) & \multicolumn{1}{l|}{1,513 (4.9\%)} & \multicolumn{1}{l|}{1,693 (4.1\%)} & \multicolumn{1}{l|}{1,596 (6.5\%)} & 2,005 (5.7\%) \\ \hline
\end{tabular}%
}
\caption{Breakdown of Characteristics by Treatment Status for Brazil Experiment. This table shows the demographics of ad viewers, both across all 24 advertising campaigns and within each of the four types of advertisements (cropped, dark complexion; uncropped, dark complexion; cropped, light complexion; and uncropped, light complexion). }
\end{table}

\begin{table}[]
\renewcommand{\tablename}{Appendix Table}
\resizebox{\columnwidth}{!}{%
\begin{tabular}{|l|l|l|l|}
\hline
 & \textbf{Dark Complexion} & \textbf{Light Complexion} & \textbf{Total} \\ \hline
Cropped & \begin{tabular}[c]{@{}l@{}}4620/30873 \\    (15.0\%)\end{tabular} & \begin{tabular}[c]{@{}l@{}}4456/24567 \\    (18.1\%)\end{tabular} & \begin{tabular}[c]{@{}l@{}}9076/55440 \\    (16.4\%)\end{tabular} \\ \hline
Uncropped & \begin{tabular}[c]{@{}l@{}}4968/40837 \\    (12.2\%)\end{tabular} & \begin{tabular}[c]{@{}l@{}}4919/34942 \\    (14.1\%)\end{tabular} & \begin{tabular}[c]{@{}l@{}}9887/75779 \\    (13.0\%)\end{tabular} \\ \hline
Total & \begin{tabular}[c]{@{}l@{}}9588/71710\\     (13.4\%)\end{tabular} & \begin{tabular}[c]{@{}l@{}}9375/59509\\     (15.8\%)\end{tabular} & \begin{tabular}[c]{@{}l@{}}18963/131219\\     (14.5\%)\end{tabular} \\ \hline
\end{tabular}%
}
\caption{Likes per Ad Views by Experimental Condition for Brazil Experiment. This figure shows the engagement rate for each experimental group. Specifically, of all advertisement Views, how many times did a viewer “Like” the picture.}
\end{table}

\begin{table}[]
\renewcommand{\arraystretch}{2}
\renewcommand{\tablename}{Appendix Table}
\resizebox{\columnwidth}{!}{%
\begin{tabular}{|l|l|l|l|l|l|}
\hline
\multicolumn{1}{|c|}{\textbf{Groups}} & \multicolumn{1}{c|}{\textbf{Image}} & \multicolumn{1}{c|}{\textbf{Likes / Views}} & \multicolumn{1}{c|}{\textbf{Image}} & \multicolumn{1}{c|}{\textbf{Likes / Views}} & \multicolumn{1}{c|}{\textbf{P-value}} \\ \hline
1 & Dark Complexion torso, uncropped & 705/4567 (15.4\%) & Light Complexion torso, uncropped & 644/4520 (14.2\%) & 0.111 \\ \hline
1 & Dark Complexion torso, cropped & 535/1783 (30\%) & Light Complexion torso, cropped & 511/1627 (31.4\%) & 0.375 \\ \hline
2 & Dark Complexion holding flowers, uncropped & 804/7263 (11.1\%) & Light Complexion holding flowers, uncropped & 873/7024 (12.4\%) & 0.011 \\ \hline
2 & Dark Complexion holding flowers, cropped & 750/7090 (10.6\%) & Light Complexion holding flowers, cropped & 850/6598 (12.9\%) & \textless{}.001 \\ \hline
3 & Dark Complexion holding hands, uncropped & 624/3543 (17.6\%) & Light Complexion holding hands, uncropped & 639/1981 (32.3\%) & \textless{}.001 \\ \hline
3 & Dark Complexion holding hands, cropped & 858/5619 (15.3\%) & Light Complexion holding hands, cropped & 658/3843 (17.1\%) & 0.158 \\ \hline
4 & Dark Complexion keyhole dress, uncropped & 1095/11650 (9.4\%) & Light Complexion keyhole dress, uncropped & 1179/10697 (11\%) & \textless{}.001 \\ \hline
4 & Dark Complexion keyhole dress, cropped & 541/3044 (17.8\%) & Light Complexion keyhole dress, cropped & 545/2576 (21.2\%) & 0.001 \\ \hline
5 & Dark Complexion racerback dress, uncropped & 992/7373 (13.5\%) & Light Complexion racerback dress, uncropped & 894/6689 (13.4\%) & 0.877 \\ \hline
5 & Dark Complexion racerback dress, cropped & 1373/10522 (13\%) & Light Complexion racerback dress, cropped & 1341/7669 (17.5\%) & \textless{}.001 \\ \hline
6 & Dark Complexion V-back dress, uncropped & 748/6441 (11.6\%) & Light Complexion V-back dress, uncropped & 690/4031 (17.1\%) & \textless{}.001 \\ \hline
6 & Dark Complexion V-back dress, cropped & 563/2815 (20\%) & Light Complexion V-back dress, cropped & 551/2254 (24.4\%) & \textless{}.001 \\ \hline
\end{tabular}%
}
\caption{Results Disaggregated by Photo for Brazil Experiment. This Table shows the engagement rate for each experimental advertisement. Specifically, of all advertisement Views, how many times did a viewer “Like” the picture.}
\end{table}


\end{document}